\documentclass[12pt]{article}
\usepackage{epsfig}
\begin{document}
\begin{titlepage}

\begin{center}

\vspace{10cm}

  {\Large \bf  Glueball dynamics in the hot plasma}
  \vspace{0.50cm}\\
  Nikolai Kochelev$^{a,b}$\footnote{Email address: kochelev@theor.jinr.ru}
\vskip 1ex

\it $^{a}$Institute of Modern Physics, Chinese Academy of Sciences, Lanzhou 730000, China
\vskip 1ex

\it $^{b}$ Bogoliubov Laboratory of Theoretical Physics, Joint
Institute for Nuclear Research, Dubna, Moscow region, 141980
Russia \vskip 1ex

\end{center}
\vskip 0.5cm \centerline{\bf Abstract}
We discuss the glueball contribution to the equation of state (EoS) of
hot gluon matter below and above $T_c$. It is shown that the strong  changing
of masses of scalar and pseudoscalar glueballs near $T_c$ plays very
important role in the thermodynamics of $SU(3)$ gauge theory.
 In particular, we give the arguments that these glueballs become massless
at $T_G\approx 1.1T_c$ and this phenomenon is crucial  in understanding
of the mystery in the behavior of trace anomaly founded in the lattice
calculation.

\end{titlepage}

\setcounter{footnote}{0}
The description of the thermodynamics of the pure $SU(3)$ gauge theory is one of the
benchmarks of the our understanding of the properties of the Quark-Gluon Plasma
(QGP). Recently a very  precise lattice results for EoS of this
theory at finite $T$ below and above deconfinement temperature $T_c$
were presented \cite{Borsanyi:2012ve}. They lead to the challenge in our understanding of the QCD dynamics at finite temperatures.
One of the puzzles is a very spectacular  behavior of the trace anomaly,
$I/T^4=(\epsilon-3p)/T^4$,
as a function of $T$. In particular, just above $T_c$ it rapidly grows till
$T_G\approx 1.1 T_c$ and then it decays  as $I/T^4\sim 1/T^2$ till $T\approx 5T_c$.
Such behaviour was found  for the first time by Pisarski by analysis of pioneer lattice calculation
in \cite{Boyd:1996bx} and was called by him as
a "fuzzy bags" manifestation \cite{Pisarski:2006yk}  (see also the discussion of this
phenomenon within the AdS/QCD approach \cite{Andreev:2007zv}).
The non-zero value of trace anomaly shows  the
 deviation of EoS from the ideal gluon gas and, therefore,
 gives the very important information about the interaction between the gluons.
 It is known that below $T_c$ the EoS for pure $SU(3)$ is described rather well by the
 gas of the massive glueballs \cite{Meyer:2009tq,Borsanyi:2012ve}. In this case the masses and
 spins of the glueballs carry  the information about interaction of the gluons
 inside glueballs (see review \cite{Mathieu:2008me}). What  can be  happened with
 glueballs  and how their properties might be changed above $T_c$ is still opened question and
 only few studies were done in this direction  \cite{Vento:2006wh,Kochelev:2006sx,Kochelev:2015rqa}.\\
In this Letter we give the arguments on the importance of the scalar and pseudoscalar glueballs
in EoS of pure SU(3) gauge theory above $T_c$. In particular, it will be shown that the most
likely scenario is that  these two glueballs
become massless  at $T_G\approx 1.1 T_c$ and this phenomenon is crucial in our understanding of
the trace anomaly behavior above  $T_c$.\\
Our starting point is the relation between lowest mass of scalar glueball, $m_G$, and gluon condensate,
$G^2=<0|\frac{\alpha_s}{\pi} G_{\mu\nu}^a G_{\mu\nu}^a|0>$ at $T=0$,
which is naturally appeared in the dilaton approach   \cite{Lanik:1984fc,Ellis:1984jv}
\begin{equation}
m_G^2f_G^2=\frac{11N_c}{6}<0|\frac{\alpha_s}{\pi}G_{\mu\nu}^a G_{\mu\nu}^a|0>,
\label{dilaton}
\end{equation}
where $f_G$ is glueball coupling constant to gluons. The lattice calculations show that gluon
condensate is decreased roughly by factor two at $T=T_c$ \cite{Lee:1989qj,DiGiacomo:1996bbx,Lee:2008xp,Miller:2006hr} due to the
strong suppression of it's electric part and slightly above $T_c$ the condensate is vanishing
very rapidly due to the cancelation between its magnetic and electric parts \cite{Lee:2008xp}. The temperature behavior of the condensate at $T_G\geq T \geq T_c$ can be described by
formula \cite{Miller:2006hr}
\begin{equation}
G^2(T)=G^2\bigg[1-\bigg(\frac{T}{T_G}\bigg)^n\bigg]
\label{cond}
\end{equation}
with $n=4$. For $T>T_G$ the gluon condensate is zero \footnote{
We should mention that the particular value of $n$ is not very important in our qualitative
discussion on the EoS below.}. If the relation Eq.\ref{dilaton} is valid also above $T_c$, then the mass
of the scalar glueball should go to zero at $T=T_G$. The possibility of the collapse $0^{++}$ glueball to
the massless state above $T_c$
due to strong attraction between two gluons in this channel induced by both the perturbative gluon exchange  and
 non-perturbative interaction with the  instanton-antiinstanton molecules,
was pointed out in  \cite{Shuryak:2004tx,Kochelev:2015rqa}.   Based on these considerations, we will assume that
 the mass of lowest  scalar glueball should be zero at $T=T_G$. We would like to emphasise that the Debye
 screening can not lead to the breakdown of the lowest mass scalar glueball state in the plasma because
 even at $T=0$ the size of this glueball is very small $R_G\leq 0.2$ fm
 \cite{Schafer:1994fd,deForcrand:1991kc,Liang:2014jta} (see discussion in \cite{Kochelev:2015rqa}).
  The important question is
 what will be happened with other  glueballs above $T_c$?  In \cite{Kochelev:2015rqa} it was argued that
   lowest pseudoscalar glueball $0^{-+}$ should be degenerate with lowest scalar glueball at $T>T_c$
   because the difference between two masses might be related to the admixture of so-called random
   instanton contribution to the vacuum state  which is strongly suppressed at $T>T_c$
   \cite{Ilgenfritz}.
   Therefore, lowest preudoscalar glueball should  become massless at $T=T_G$ as well. It is evident, that just after
   $T_c$ all others heavy glueballs should decay very fast to these light glueballs.
   Therefore, we come to conclusion that thermodynamics of the theory at  $T_G\geq T \geq T_c$
   is determined by these light glueballs only.
   In this phase the gluons are still confined inside two glueballs and  pressure is
   \begin{equation}
   P(T))_{T_G>T>T_c}=-T\frac{N_G}{2\pi^2}\int_0^\infty k^2dk Log\bigg[1-exp\bigg(-\frac{\sqrt{k^2+m^2(T)}}{T}\bigg)\bigg],
   \label{pressure}
   \end{equation}
where $N_G=2$ for two glueball states. The trace anomaly  for the case of the temperature's dependent mass of the boson in hot plasma is given by formula \cite{castorina} 
\begin{eqnarray}
\frac{I(T)}{T^4}=T\frac{d}{dT}\bigg(\frac{P}{T^4}\bigg)&=&\frac{N_G}{2\pi^2}\int_0^\infty dx\frac{x^2}{\sqrt{x^2+m(T)^2/T^2}
(e^{\sqrt{x^2+m(T)^2/T^2}}-1)}\nonumber\\
&\times &
\frac{m(T)^2}{T^2}
\bigg[1-\frac{T}{m(T)}\frac{dm(T)}{dT}\bigg],
\label{trace1}
\end{eqnarray}
where according with Eq.\ref{dilaton} and Eq.\ref{cond}
\begin{equation}
m(T)=m_0\sqrt{1-\bigg(\frac{T}{T_G}\bigg)^4}.
\label{mass}
\end{equation}
Above $T_G$ the massless glueballs can dissociate to the massless gluons due to the reaction
$G+G\rightarrow gluon+gluon$. Therefore, at $T>T_G$ it should be  mixed glueball-gluon phase.
At very large temperature we should have free gluon gas only and the contribution
of glueballs to the pressure should  be suppressed to respect to free gluon case by
factor $(T_G/T)^\lambda $  with $\lambda>0$. Based on the scaling in trace anomaly
$I(T)/T^4(T/T_c)^2$ observed in
lattice calculation \cite{Borsanyi:2012ve} we will use $\lambda=2$.
With this assumption the pressure above $T_G$ is the sum of glueball and gluon pressures
\begin{equation}
P(T)_{T>T_G}=\frac{N_G\pi^2}{90}T^4\bigg(\frac{T_G}{T}\bigg)^2+\frac{N_g\pi^2}{90}T^4
\bigg[1-\bigg(\frac{T_G}{T}\bigg)^2\bigg],
\label{pressure2}
\end{equation}
where $N_g=2(N_c^2-1)=16$ is the number of the gluon degrees of freedom.
It is natural to assume also that  $T_G$ is related to the value of the temperature at
 which the maximum of trace anomaly was observed. Therefore, we will put $T_G\approx 1.1T_c$.
 In this case for the trace anomaly at $T=T_G$, Eq.\ref{pressure2},  gives
 $\Delta_{max}=I(T_G)/T_G^4=(N_g-N_G)\pi^2/45=3.07$  which is in the good agreement with the  lattice
 result $\Delta_{max}^{lattice}=2.48$ \cite{Borsanyi:2012ve}.
 Futhermore, the matching trace anomaly values in light glueball and glueball-gluon phases
 at $T=T_G$ gives the relation between parameter $m_0$ in glueball mass formula and $T_G$
 \begin{equation}
 m_0=\sqrt{3\Delta_{max}}T_G.
 \label{m0}
 \end{equation}
 For the $T_c\approx 270$ MeV in pure $SU(3)$ gauge theory we obtain
 $m_0=901$ MeV and $m(T_c)=507$ MeV. It means that indeed in the temperature's range
  $T_G>T>T_c$ we have
 very light scalar and pseudiscalar glueballs, $m(T_c)>m(T)>0$.
 \begin{figure}[h]
\begin{minipage}[c]{6cm}
\hspace*{1.0cm}
\centerline{\epsfig{file=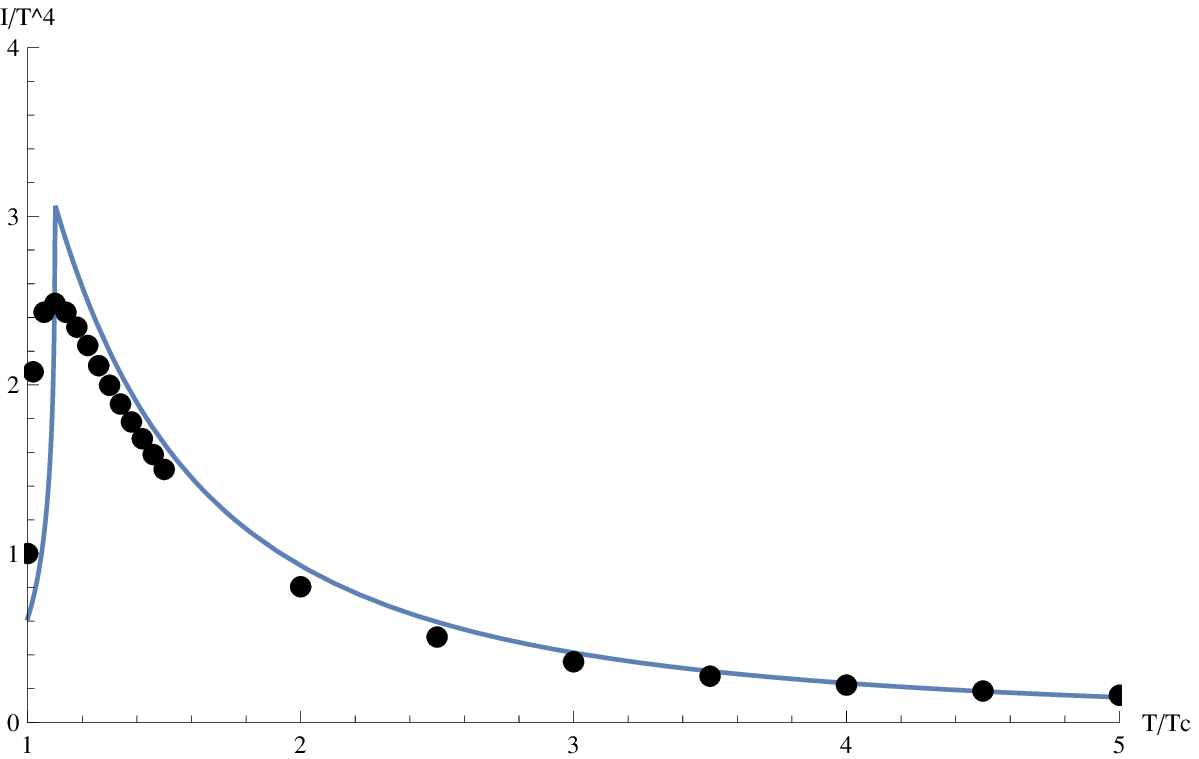,width=6cm,height=5cm, angle=0}}\
\end{minipage}
\begin{minipage}[c]{6cm}
 \hspace*{3.0cm}
\centerline{\epsfig{file=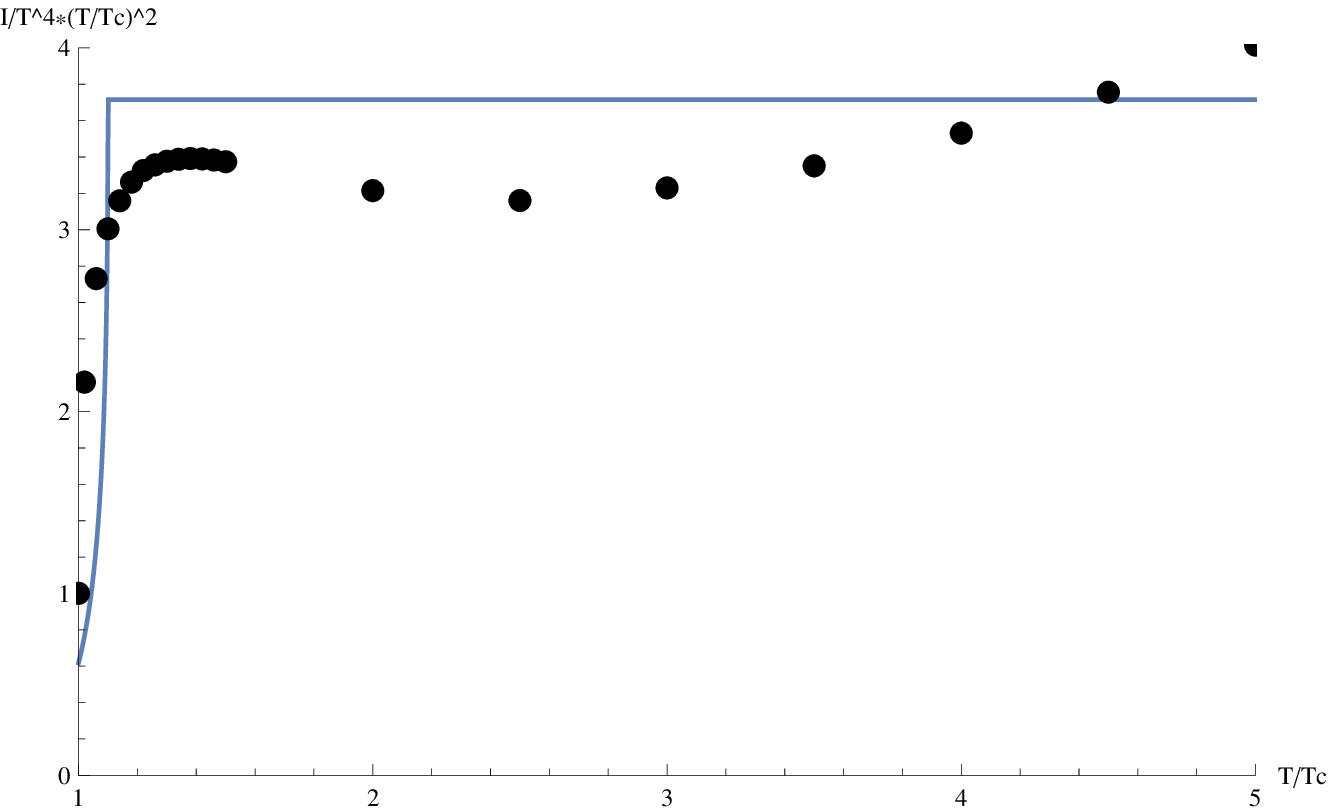,width=6cm,height=5cm, angle=0}}\
\end{minipage}
\caption{ The trace anomaly $I(T)/T^4$ (left panel), and its  scaling value $(I(T)/T^4)(T/T_c)^2$
 (right panel) as the function of $T/T_c$. The lattice data
  presented by the bold spots \cite{Borsanyi:2012ve}.}
\end{figure}
 \begin{figure}[h]
\centerline{\epsfig{file=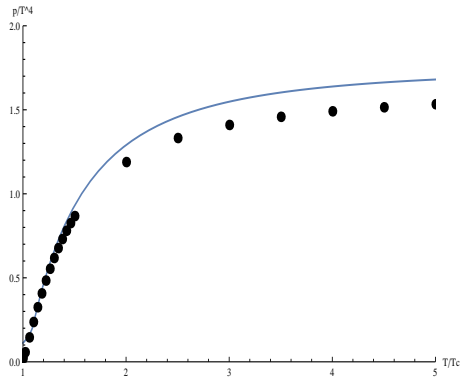,width=6cm,height=5cm, angle=0}}\
\caption{ The pressure $p/T^4$ as the function of $T/T_c$  in comparison with the  lattice data \cite{Borsanyi:2012ve}.}
\end{figure}
 In the Figs.1,2 the result of the trace anomaly and pressure  calculation is presented. One can see
 they are  in the qualitative agreement with the lattice data
 at the $5T_c\geq T\geq T_c $ \cite{Borsanyi:2012ve}. We would like to  emphasize that our simple model
 does not include perturbative QCD corrections to EoS. Therefore, it might be possible to
 improve the agreement of the model  with the data after the consideration of such corrections.
  It should be also mentioned that our model at $T>T_G$, Eq.\ref{pressure2},  predicts  the small violation of the $2(N_c^2-1)$
 scaling for the trace anomaly observed in the  lattice calculation for the large $N_c$ case \cite{Panero:2009tv}. \footnote{The author
 is grateful to Oleg Andreev which attracted his attention to this phenomenon.} It would be interesting
 to check this prediction with a more precise lattice calculation.
 \\
 In summary, we can conclude that  there are three phases in the pure $SU(3)$ gauge theory at finite
 $T$. At $T<T_c$ there is the gas of the very massive glueballs. Just above $T_c$ at  $T_G\geq T\geq T_c $
 the phase with the light scalar and pseudoscalar mesons is dominated. After $T_G\approx 1.1T_c$
 we have mixed massless gluons and scalar-pseudoscalar massless  glueballs phase.
It is evident that this phenomenon should also give the influence to the EoS of the full QCD with the
light quarks.
In this case, the mixing between the glueballs and the quark-antiquark states should be considered as well.
The study of these effects is the subject of our future investigation.

\section{Acknowledgments}

I would like to thank  Adriano Di Giacomo, Oleg Andreev, Ernst-Michael Ilgenfritz, Robert Pisarski and  Vicente Vento
   for useful
discussions.
I deeply appreciate the  warm hospitality by  Pengming Zhang at the  Institute of Modern Physics, CAS in Lanzhou,
where this work was completed.
This work was partially supported by the
National Natural Science Foundation of China (Grant No. 11035006 and 11175215),
and by the Chinese
Academy of Sciences visiting professorship for senior international scientists (Grant No. 2013T2J0011).
The work was initiated through the series of APCTP-BLTP JINR Joint Workshops.

\end{document}